\documentclass[twocolumn,superscriptaddress,preprintnumbers,amsmath,amssymb,pra]{revtex4-2}

\usepackage{graphicx}
\usepackage{dcolumn}
\usepackage{bm}
\usepackage{ifpdf}
\usepackage{epstopdf}
\usepackage{slashed}
\usepackage{amsfonts}
\usepackage{mathrsfs}
\usepackage{amssymb}
\usepackage{multirow}
\usepackage{CJK}
\usepackage{xcolor}
\usepackage{textcomp}
\usepackage{csquotes}
\usepackage{tikz}
\usepackage{tikz-feynman}
\tikzfeynmanset{compat=1.1.0}

\def\lesssim{\ \raise.3ex\hbox{$<$}\kern-0.8em\lower.7ex\hbox{$\sim$}\ }
\def\gesim{\ \raise.3ex\hbox{$>$}\kern-0.8em\lower.7ex\hbox{$\sim$}\ }

\begin{document}
\begin{CJK}{UTF8}{ipxm}

\title{Growth of quartet correlations in neutron-rich tellurium isotopes within quartet Bardeen-Cooper-Schrieffer theory}

\author{Yixin Guo (郭一昕)}
\email{yixin.guo@riken.jp}
\affiliation{RIKEN Nishina Center for Accelerator-Based Science, Wako 351-0198, Japan}

\author{Hiroyuki Sagawa (佐川弘幸)}
\email{sagawa@ribf.riken.jp}
\affiliation{RIKEN Nishina Center for Accelerator-Based Science, Wako 351-0198, Japan}
\affiliation{Center for Mathematics and Physics, University of Aizu, Aizu Wakamatsu, Fukushima 965-0001, Japan}
\affiliation{Institute of Theoretical Physics, Chinese Academy of Sciences, Beijing 100190, China}

\author{Masaaki Kimura (木村真明)}
\email{masaaki.kimura@ribf.riken.jp}
\affiliation{RIKEN Nishina Center for Accelerator-Based Science, Wako 351-0198, Japan}
\affiliation{Department of Physics, Graduate School of Science, The University of Tokyo, Tokyo 113-0033, Japan}
\affiliation{Quark Nuclear Science Institute, The University of Tokyo, Tokyo 113-0033, Japan}

\date{\today}

\begin{abstract}
Quartet correlations in neutron-rich Te isotopes are investigated within the quartet Bardeen-Cooper-Schrieffer (BCS) framework.  
Taking \(^{100}\)Sn as an inert core, we consider two valence protons and valence neutrons occupying the \(2d_{5/2}\oplus1g_{7/2}\) model space, and solve the quartet BCS variational equations with a charge-independent isovector pairing interaction.  
The effective pairing strength is constrained from empirical neutron pairing gaps in the Te isotopic chain.  
We find that the valence quartet number increases as the valence neutron number is enlarged from \(N_{\rm val}=2\) to \(14\).  
The same increasing behavior is also found for the condensed quartet component.  
The proton occupation of the \(1g_{7/2}\) orbit is strongly enhanced relative to the conventional like-particle BCS reference and is driven close to the degeneracy-weighted limit.  
These results suggest that additional valence neutrons enhance the quartet admixture in the correlated quartet BCS state, while redistributing the fixed proton weight from pair-like configurations to quartet configurations.
\end{abstract}

\maketitle

\section{Introduction}\label{sec:I}

Clustering is one of the characteristic manifestations of strong many-body correlations in atomic nuclei~\cite{vonOertzen2006Phys.Rep.432.43113, Freer2018Rev.Mod.Phys.90.035004}.  
Among various nuclear clusters, the $\alpha$ particle is particularly important because of its large binding energy and its spin-isospin saturated structure.  
The formation of $\alpha$-like four-nucleon correlations has long been discussed in self-conjugate light nuclei, dilute nuclear matter, and nuclear surfaces, where it is closely related to the interplay between pairing correlations, Pauli blocking, and the low-density environment of finite nuclei~\cite{Schuck2007Prog.Part.Nucl.Phys.59.285--304, Tohsaki2001Phys.Rev.Lett.87.192501,Funaki2008Phys.Rev.C77.064312}.  
In particular, microscopic studies of dilute nuclear matter have shown that the formation and dissolution of $\alpha$-like correlations are governed by medium effects and Pauli blocking~\cite{Roepke1998Phys.Rev.Lett.80.3177--3180, Beyer2000Phys.Lett.B488.247--253}.  
Such correlations are also relevant to astrophysical environments, where light clusters can appear in dilute matter and affect the nuclear equation of state~\cite{Typel2010Phys.Rev.C81.015803,Oertel2017Rev.Mod.Phys.89.015007}.

The conventional Bardeen-Cooper-Schrieffer (BCS) theory and its nuclear extensions provide a successful framework for describing two-body pairing correlations in nuclei~\cite{Bardeen1957Phys.Rev.108.1175--1204,Bohr1958Phys.Rev.110.936938,peter}.  
Nuclear pairing~\cite{Dean2003Rev.Mod.Phys.75.607--656} is reflected, for example, in the odd-even staggering of binding energies and plays a central role in open-shell nuclei.  
However, $\alpha$-like correlations involve two neutrons and two protons and therefore go beyond the ordinary two-body pairing picture~\cite{Frauendorf2014Prog.Part.Nucl.Phys.78.2490}.  
In order to describe such four-body correlations, {particle-number-projected variational wave functions that treat both $T=1$ and $T=0$ neutron-proton pairing and explicitly include $\alpha$-like correlations have been developed~\cite{Chasman2002PhysLettB524.81,Chasman2003PhysLettB553.204}.  
These calculations focused on nuclei on or close to the $N=Z$ line and showed that the quartet component can be substantial when the proton-neutron balance and single-particle structure are favorable.  
Related studies examined the contributions of diagonal neutron-proton pairing matrix elements to the Wigner and symmetry energies and to shell gaps~\cite{Chasman2003PhysLettB577.47,Chasman2007PhysRevLett99.082501}.  
Together, these works provide important precedents for variational treatments of pair and quartet correlations in finite nuclei.  
Subsequent approaches include quartet-condensation models~\cite{Sandulescu2012Phys.Rev.C85.061303,Sambataro2015Phys.Rev.Lett.115.112501,Negrea2018Phys.Rev.C98.064319} and quartet extensions of BCS-type variational states~\cite{Baran2020Phys.Lett.B805.135462,Baran2020Phys.Rev.C102.061301}.  
These studies show that quartet correlations can coexist with, and in some cases compete with, ordinary pair correlations.}

A full quartet BCS framework provides a natural extension of the BCS idea from coherent pair amplitudes to coherent quartet amplitudes. 
Although the quartet BCS wave function does not conserve particle number exactly, it offers a transparent variational framework in which pair-like and quartet-like components can be treated on the same footing.
The quartet BCS theory has been applied to infinite symmetric nuclear matter, where Cooper quartet correlations and the coexistence of pair and quartet condensates were investigated \cite{Guo2022Phys.Rev.C105.024317,Guo2022Phys.Rev.Research4.023152}.  
Combined with a local-density description, the quartet BCS theory has also been used to discuss quartet correlations near the surface of $N=Z$ doubly magic nuclei~\cite{Guo2025Phys.Rev.C112.024310}.  
Moreover, the validity of quartet BCS theory has been examined via a comparative study of the quartet superfluid state with generalized Nambu-Gor'kov formalism~\cite{Sogo2010Phys.Rev.C81.064310,guo2026comparativestudyquartetsuperfluid}.
These works suggest that quartet BCS theory is a useful tool for exploring four-body correlations beyond the conventional pairing paradigm.

The region around the doubly magic $^{100}$Sn core provides a particularly attractive testing ground for $\alpha$-like correlations~\cite{Hinke2012Nature486.341345}.
The enhanced $\alpha$-decay expected above $^{100}$Sn was already discussed in connection with superallowed $\alpha$ decay~\cite{Macfarlane1965Phys.Rev.Lett.14.114115}. 
Subsequent experimental studies identified $\alpha$-emitting nuclei in this region, including $^{105}$Te and $^{109}$Xe~\cite{Liddick2006Phys.Rev.Lett.97.082501}, and the decay chain $^{108}$Xe$\rightarrow^{104}$Te$\rightarrow^{100}$Sn~\cite{Auranen2018Phys.Rev.Lett.121.182501}. 
These observations, together with microscopic and cluster-model calculations of $^{104}$Te and neighboring nuclei~\cite{Mohr2007Eur.Phys.J.A31.2328,Patial2016Phys.Rev.C93.054326,Yang2020Phys.Rev.C101.024316}, suggest that the $^{100}$Sn region is sensitive to the interplay between shell closure, proton-neutron correlations, and $\alpha$-like quartet formation.
More recently, quasifree $(p,p\alpha)$ reaction measurements on Sn isotopes have probed surface $\alpha$-cluster formation and its neutron-number dependence~\cite{Tanaka2021Science371.260264}.

Motivated by these experimental and theoretical studies
of $\alpha$-like correlations, we apply the quartet BCS framework to neutron-rich Te isotopes outside the $^{100}$Sn core.  
We consider two valence protons and valence neutrons occupying the $2d_{5/2}\oplus1g_{7/2}$ shell space and use a charge-independent isovector pairing interaction. The effective pairing strength is constrained from empirical neutron pairing gaps in the Te isotopic chain.  
Within this framework, we find that the valence quartet number grows as the valence neutron number is increased from $N_{\rm val}=2$ to $N_{\rm val}=14$. The corresponding quartet-induced energy gain shows the same trend. 
We also show that, as neutrons are added, the two valence protons are more strongly involved in quartet configurations.

This paper is organized as follows.  
In Sec.~\ref{sec:II}, we formulate the quartet BCS framework for finite nuclei in the \(2d_{5/2}\oplus1g_{7/2}\) valence space outside the \(^{100}\)Sn core.  
In Sec.~\ref{sec:III}, we present the numerical results for Te  isotopes, including the valence quartet number, the quartet-induced energy gain, the proton occupation of the \(1g_{7/2}\) orbit, and the proton gap-like quantity.  
Finally, a summary and perspectives are given in Sec.~\ref{sec:IV}.

\section{Theoretical framework}\label{sec:II}
We consider a finite valence system of neutrons and protons outside a
self-conjugate inert core, interacting through charge-independent
pairing forces that scatter pairs of nucleons between time-reversed
single-particle states. 
We label a
single-particle state by $i\equiv(a_i,m_i)$ with
$a_i=(n_i l_i j_i)$ and $\bar i\equiv(a_i,-m_i)$ denoting the
time-reversed partner.
The notation $i>0$ indicates that one state from each time-reversed
pair is included in the sum.
We use $c_{i\nu}\equiv\nu_i$ and $c_{i\pi}\equiv\pi_i$.
The Hamiltonian is written as
\begin{align}\label{hamiltonian}
H =\,&
\sum_{i>0}\sum_{\tau=\nu,\pi}
\varepsilon_{a_i,\tau}
\left(
c^\dagger_{i\tau}c_{i\tau}
+c^\dagger_{\bar i\tau}c_{\bar i\tau}
\right)\nonumber\\
&+
\sum_{T_3}\sum_{i>0}\sum_{i'>0}
\widetilde V_{a_i a_{i'}}
P^\dagger_{i,T_3}P_{i',T_3}.
\end{align}
The reduced isovector interaction strength $\widetilde V_{a_i a_{i'}}$
and the pair operators entering Eq.~\eqref{hamiltonian} are
defined as
\begin{align}
\widetilde V_{a_i a_{i'}}
&=
\frac{V_{a_i a_{i'}}}
{\sqrt{2j_i+1}\sqrt{2j_{i'}+1}},\\
P^\dagger_{i,+1}&=(-1)^{j_i-m_i}
\nu_i^\dagger\nu_{\bar i}^\dagger,\\
P^\dagger_{i,0}&=(-1)^{j_i-m_i}
\frac{1}{\sqrt{2}}
\left(
\nu_i^\dagger\pi_{\bar i}^\dagger
+\pi_i^\dagger\nu_{\bar i}^\dagger
\right),\\
P^\dagger_{i,-1}&=(-1)^{j_i-m_i}
\pi_i^\dagger\pi_{\bar i}^\dagger .\label{eq:5}
\end{align}
Here, we follow the convention that the third component of isospin for single nucleon is $+\frac{1}{2}$ for neutrons and $-\frac{1}{2}$ for protons.
Equations~\eqref{hamiltonian}--\eqref{eq:5} define the Hamiltonian used below. 
In the present first-step treatment we retain only time-reversed isovector $J=0$, $T=1$ pairs within the same spherical orbit. 
This is the most established and dominant pairing mode in finite nuclei~\cite{Dean2003Rev.Mod.Phys.75.607--656,Frauendorf2014Prog.Part.Nucl.Phys.78.2490}.
Possible isoscalar proton-neutron pairing correlations are not included explicitly. 
While such correlations may affect the absolute quartet amplitudes and correlation energies, especially near $N=Z$, they are not expected to alter the qualitative neutron-number trend found here. 
Their inclusion is left for future work.
Cross-orbital pair operators, such as $d_{5/2}$--$g_{7/2}$ pairs, are not included explicitly, while interorbital coupling enters through the pair-scattering matrix elements $\widetilde V_{a_i a_{i'}}$.

For the numerical application to Te isotopes, we take $^{100}$Sn as an inert core and fix the number of valence protons to two, allowing them to occupy both the $2d_{5/2}$ and $1g_{7/2}$ orbitals. 
The valence neutrons are also allowed to occupy both orbitals. 
Following the experimental assignment~\cite{Seweryniak2007Phys.Rev.Lett.99.022504}, the same relative single-particle spacing is used for neutrons and protons,
$\varepsilon_{2d_{5/2},\nu}=\varepsilon_{2d_{5/2},\pi}=0$ and
$\varepsilon_{1g_{7/2},\nu}=\varepsilon_{1g_{7/2},\pi}=0.172~{\rm MeV}$.
A constant shift between the neutron and proton single-particle spectra can be absorbed into the corresponding chemical potentials and does not affect the variational amplitudes at fixed neutron and proton numbers.
In this way, the role of different-$j$ neutron configurations can be examined while keeping the proton sector unchanged.

Although an $\alpha$-like quartet with total angular momentum
$J=0$ does not in general imply that every two-body subsystem is
coupled to $J=0$, quartet-based studies indicate that the low-energy
structure of self-conjugate nuclei is often dominated by $J=0$,
$T=0$ four-body correlations~\cite{Sambataro2015Phys.Rev.Lett.115.112501}. 
At the same time, the dominant pairing collectivity in finite nuclei is commonly associated with isovector $J=0$, $T=1$ pairing components~\cite{Dean2003Rev.Mod.Phys.75.607--656,Frauendorf2014Prog.Part.Nucl.Phys.78.2490}. 
Motivated by this picture, we now construct the quartet BCS trial state from the time-reversed isovector pairs defined above.

For compactness, we relabel the single-particle index entering the
time-reversed pair operator as $r\equiv i=(a_r,m_r)$, and denote a
two-pair block by
$b\equiv(r_1,r_2)$ with $r_1\le r_2$.
Here, a ``block'' refers to the empty, pair-like, and quartet configurations associated with a fixed pair of labels $(r_1,r_2)$, which together form one factor of the product trial state in Eq.~\eqref{twf}.
The quartet creation operator is then introduced as
\begin{align}
\alpha^\dagger_{r_1r_2}
=\,&
\frac{1}{\sqrt{3}}\frac{1}{\sqrt{1+\delta_{r_1r_2}}}\nonumber\\
\,&\times\left(
P^\dagger_{r_1,+1}P^\dagger_{r_2,-1}
+P^\dagger_{r_1,-1}P^\dagger_{r_2,+1}
-P^\dagger_{r_1,0}P^\dagger_{r_2,0}
\right),
\end{align}
and the trial wave function is constructed as
\begin{align}\label{twf}
|\Psi\rangle
=\,&
\prod_{r_1\le r_2} |\Psi_{r_1r_2}\rangle,\nonumber\\
|\Psi_{r_1r_2}\rangle
=\,&
\left[
u_{r_1r_2}
+\sum_{T_3}v_{r_1r_2;T_3}B^\dagger_{r_1r_2;T_3}
+w_{r_1r_2}\alpha^\dagger_{r_1r_2}
\right]|0\rangle,
\end{align}
where
\(|0\rangle\) denotes the particle vacuum with respect to the inert
core and
\begin{align}
B^\dagger_{r_1r_2;T_3}
=
\frac{
P^\dagger_{r_1,T_3}+P^\dagger_{r_2,T_3}
}
{\sqrt{2(1+\delta_{r_1r_2})}}.
\end{align}
The normalization condition
$\left\langle\Psi_{r_1r_2}|\Psi_{r_1r_2}\right\rangle=1$ gives
\begin{align}\label{norm}
|u_{r_1r_2}|^2+\sum_{T_3}|v_{r_1r_2;T_3}|^2
+|w_{r_1r_2}|^2=1 .
\end{align}
{In the present work, we employ a finite-nucleus quartet BCS construction tailored to neutron-rich Te isotopes.  
The trial wave function in Eq.~\eqref{twf} is a product of two-pair blocks, each containing empty, pair-like, and quartet-like amplitudes, while the neutron and proton numbers are constrained on average.  
Such a block-product ansatz differs from the particle-number-projected variational wave functions used in Refs.~\cite{Chasman2002PhysLettB524.81,Chasman2003PhysLettB553.204}.  
We apply it in a specified $2d_{5/2}\oplus1g_{7/2}$ model space to examine how quartet correlations evolve with neutron number while the average number of active protons is fixed at two.}  

In what follows, we use the compact notation
$u_b\equiv u_{r_1r_2}$, $v_{b,T_3}\equiv v_{r_1r_2;T_3}$, and
$w_b\equiv w_{r_1r_2}$ for the variational amplitudes in block $b$.
The present two-orbit model space contains seven time-reversed pair
labels $r$, and therefore 28 two-pair blocks $b=(r_1,r_2)$.
For later use, we introduce the block-dependent quantities
\begin{align}
\delta_b&\equiv\delta_{r_1r_2},\\
M_r^{(b)}
&\equiv
\delta_{r r_1}
+
\delta_{r r_2},\\
\eta_r^{(b)}
&\equiv
\frac12 M_r^{(b)},\\
\mathcal{A}_r^{(b)}
&\equiv
\frac{M_r^{(b)}}{\sqrt{2\left(1+\delta_b\right)}},
\\
\mathcal{B}_r^{(b)}
&\equiv
\frac{\left(M_r^{(b)}\right)^2}{\sqrt{6}\left(1+\delta_b\right)},
\\
\varepsilon_{b,\tau}
&\equiv
\sum_r\eta_r^{(b)}\varepsilon_{a_r,\tau}.
\end{align}
Here, $M_r^{(b)}$ counts how many times the pair label $r$ appears in the block $b$ and $\eta_r^{(b)}$ is the corresponding weight entering the block-averaged single-particle energy.
We also define the isospin-dependent sign factor
$\sigma_{+1}=\sigma_{-1}=+1$ and $\sigma_0=-1$.

Using the trial wave function~\eqref{twf}, the single-particle
contribution to the energy is
\begin{align}
\langle H_{\rm sp}\rangle
=
\sum_b
\left[
\sum_{T_3}|v_{b,T_3}|^2\,
E^{(B)}_{b,T_3}
+
|w_b|^2\,
E^{(\alpha)}_{b}
\right],
\end{align}
while the pairing contribution can be written as
\begin{align}
\langle H_{\rm pair}\rangle
=
\sum_{T_3}\sum_{r,s}
\widetilde V_{a_r a_s}\,
\kappa_{r,T_3}^*\,
\kappa_{s,T_3},
\end{align}
\begin{align}
&\kappa_{r,T_3}
\equiv
\langle  P_{r,T_3}\rangle\nonumber\\
=\,&
\sum_b
\Biggl\{
\frac{
M_r^{(b)}
}{
\sqrt{2\left(1+\delta_b\right)}
}
\,u_b^*\,
v_{b,T_3}
\nonumber\\
&+
\sigma_{T_3}\,
\frac{
\left[M_r^{(b)}\right]^2
}{\sqrt{6}
\left(1+\delta_b\right)
}
\,w_b\,
v_{b,-T_3}^*
\Biggr\},
\end{align}
where
\begin{align}
E^{(B)}_{b,T_3}
=\,&
\frac12
\left(
E^{(P)}_{r_1,T_3}
+
E^{(P)}_{r_2,T_3}
\right),
\\
E^{(P)}_{r,T_3}
=\,&
\begin{cases}
2\varepsilon_{a_r,\nu}, & T_3=+1,\\
\varepsilon_{a_r,\nu}+\varepsilon_{a_r,\pi}, & T_3=0,\\
2\varepsilon_{a_r,\pi}, & T_3=-1,
\end{cases}\\
E^{(\alpha)}_{b}
=\,&
\varepsilon_{a_{r_1},\pi}+\varepsilon_{a_{r_1},\nu}
+\varepsilon_{a_{r_2},\pi}+\varepsilon_{a_{r_2},\nu}
\end{align}
are introduced.

Combining the two contributions gives the expectation value of the full Hamiltonian
\begin{align}
&\langle \Psi|H|\Psi\rangle\nonumber\\
=\,&
\sum_{r}\sum_{b}\eta_r^{(b)}
\Big[
2\varepsilon_{a_r,\nu}|v_{b,+1}|^2
+
(\varepsilon_{a_r,\nu}+\varepsilon_{a_r,\pi})|v_{b,0}|^2\nonumber\\
&+
2\varepsilon_{a_r,\pi}|v_{b,-1}|^2
+
2(\varepsilon_{a_r,\nu}+\varepsilon_{a_r,\pi})|w_b|^2
\Big]\nonumber
\\
&+
\sum_{T_3}\sum_{r,s}
\widetilde V_{a_r a_s}\,
\kappa_{r,T_3}^*
\kappa_{s,T_3},
\end{align}
where the first term is the explicit single-particle contribution and
the second term is the pairing contribution expressed through
\(\kappa_{r,T_3}\).

Because a common phase rotation of all amplitudes \((u_b,v_{b,T_3},w_b)\) changes the block state only by an overall phase, we use this freedom to choose \(u_b\) real, while retaining \(v_{b,T_3}\) and \(w_b\) as complex variables so that the relative phases entering \(\kappa_{r,T_3}\) remain variational degrees of freedom.
With this phase convention, the normalization condition~\eqref{norm} gives its variation with respect to \(v_{b,T_3}^*\) and \(w_b^*\) as
\begin{align}
\delta u_b
=
-\frac{1}{2u_b}
\left(
\sum_{T_3} v_{b,T_3}\,\delta v_{b,T_3}^*
+
w_b\,\delta w_b^*
\right).
\end{align}

We further introduce the order parameter
\begin{align}
\Delta_{s,T_3}
=
-\sum_{r}
\widetilde V_{a_s a_r}
\sum_{b}
\eta_r^{(b)}
\Big(
\mathcal{A}_r^{(b)}u_b v_{b,T_3}
+\sigma_{T_3}\mathcal{B}_r^{(b)}w_b v_{b,-T_3}^*
\Big)
\end{align}
and
\begin{align}
B_b
=
\frac{4}{u_b}
\sum_{T_3}\sum_{r,s}
\eta_r^{(b)}
\mathrm{Re}
\left(
\mathcal{A}_r^{(b)} v_{b,T_3}^* \Delta_{s,T_3}
\right).
\end{align}
The valence neutron and proton numbers are constrained by minimizing
the Routhian
\begin{align}
\mathcal{R}
=
\langle \Psi|H|\Psi\rangle
-\mu_\nu N_\nu
-\mu_\pi N_\pi ,
\end{align}
where \(\mu_\nu\) and \(\mu_\pi\) are the neutron and proton chemical
potentials, respectively.  Equivalently, the
single-particle block energies entering the denominators below are
shifted to
\begin{align}
\xi_{b,\tau}
\equiv
\varepsilon_{b,\tau}-\mu_\tau,
\qquad \tau=\nu,\pi .
\end{align}
With these definitions, the resulting variational equations are
\begin{align}
v_{b,T_3}
=
\frac{
4\sum_{r,s}\eta_r^{(b)}
\left(
\mathcal{A}_r^{(b)}u_b\Delta_{s,T_3}
+\sigma_{T_3}
\mathcal{B}_r^{(b)}w_b\Delta_{s,-T_3}^*
\right)
}{
D_{b,T_3}
},
\end{align}
\begin{align}
w_b
=
\frac{
4\sum_{T_3}\sum_{r,s}
\eta_r^{(b)}
\sigma_{T_3}v_{b,-T_3}\Delta_{s,T_3}
}{
D_{b,Q}
},
\end{align}
where
\begin{align}
D_{b,+1}&\equiv B_b+2\xi_{b,\nu},
\quad
D_{b,0}\equiv B_b+\xi_{b,\nu}+\xi_{b,\pi},
\nonumber\\
D_{b,-1}&\equiv B_b+2\xi_{b,\pi},
\quad
D_{b,Q}\equiv B_b+2\xi_{b,\nu}+2\xi_{b,\pi}.
\end{align}

The particle numbers are given by
\begin{align}
N_\nu
&
=
2\sum_{b}
\left(
|v_{b,+1}|^2
+\frac12|v_{b,0}|^2
+|w_{b}|^2
\right),
\\
N_\pi
&
=
2\sum_{b}
\left(
|v_{b,-1}|^2
+\frac12|v_{b,0}|^2
+|w_{b}|^2
\right).
\end{align}
Although the quartet BCS state is not an eigenstate of the particle-number
operators, the desired valence neutron and proton numbers are imposed
on their expectation values.  In practice, \(\mu_\nu\) and
\(\mu_\pi\) are solved self-consistently together with the gap-like
quantities so that \(N_\nu=N_\nu^{\rm target}\) and
\(N_\pi=N_\pi^{\rm target}\).

\section{Results and discussion}\label{sec:III}

In order to investigate quartet correlations in neutron-rich tellurium isotopes, we adopt the constant-$G$ approximation~\cite{peter} and
consider spherical nuclei without deformation.  We determine the
effective pairing strength \(G\) from the even-even Te isotopes away
from the proton dripline, where the empirical pairing gaps are expected
to be less affected by weak binding and continuum effects.  The more
proton-rich nuclei are then treated as an extrapolation of the same
finite-space interaction toward the dripline.  In the constant-\(G\)
approximation,
\begin{align}
    \widetilde V_{aa'}\equiv\frac{G}{\sqrt{(2j_a+1)(2j_{a'}+1)}}
\end{align}
and there are only two kinds of order parameters, namely, $\Delta_{d,T_3}$ and $\Delta_{g,T_3}$.
Moreover, in order to measure the nucleon pairing scale in the present multiple $j$ framework, we define a degeneracy-weighted sum of orbital gaps as
\begin{align}
\Xi_{T_3}\sum_j \Omega_j
\equiv
\sum_{s}\Delta_{s,T_3}
=
\sum_j \Omega_j\Delta_{j,T_3}
\quad\mbox{with}\quad
\Omega_j\equiv\frac{2j+1}{2}.
\end{align}
In the present $d_{5/2}\oplus g_{7/2}$ space, this means
\begin{align}
7\Xi_{T_3}=3\Delta_{d,T_3}+4\Delta_{g,T_3}.
\end{align}
As a practical criterion for fixing the neutron pairing strength $G$, we use the empirical neutron pairing scale extracted from odd-even mass staggering. 
For even-$N$ nuclei, we adopt the standard three-point indicator~\cite{nsbohr,Satula1998Phys.Rev.Lett.81.35993602,Changizi2015Nucl.Phys.A940.210226}
\begin{align}
&\Delta_n^{(3)}(N,Z)
=
\frac{(-1)^{N+1}}{2}\nonumber\\
&\qquad\times\Bigl[
B(N-1,Z)+B(N+1,Z)-2B(N,Z)
\Bigr],
\end{align}
where $B(N,Z)$ is the nuclear binding energy. 
Following the analysis of Ref.~\cite{Changizi2015Nucl.Phys.A940.210226}, this quantity provides a reasonable proxy for the neutron pairing gap in even-$N$ systems. 
Using AME2020 masses~\cite{Wang_2021} as the experimental mass standard, the Te  isotopic chain in the region considered here suggests an empirical neutron pairing scale of order \(1.2\text{--}1.4\) MeV. 

\begin{figure}
\includegraphics[width=\columnwidth]{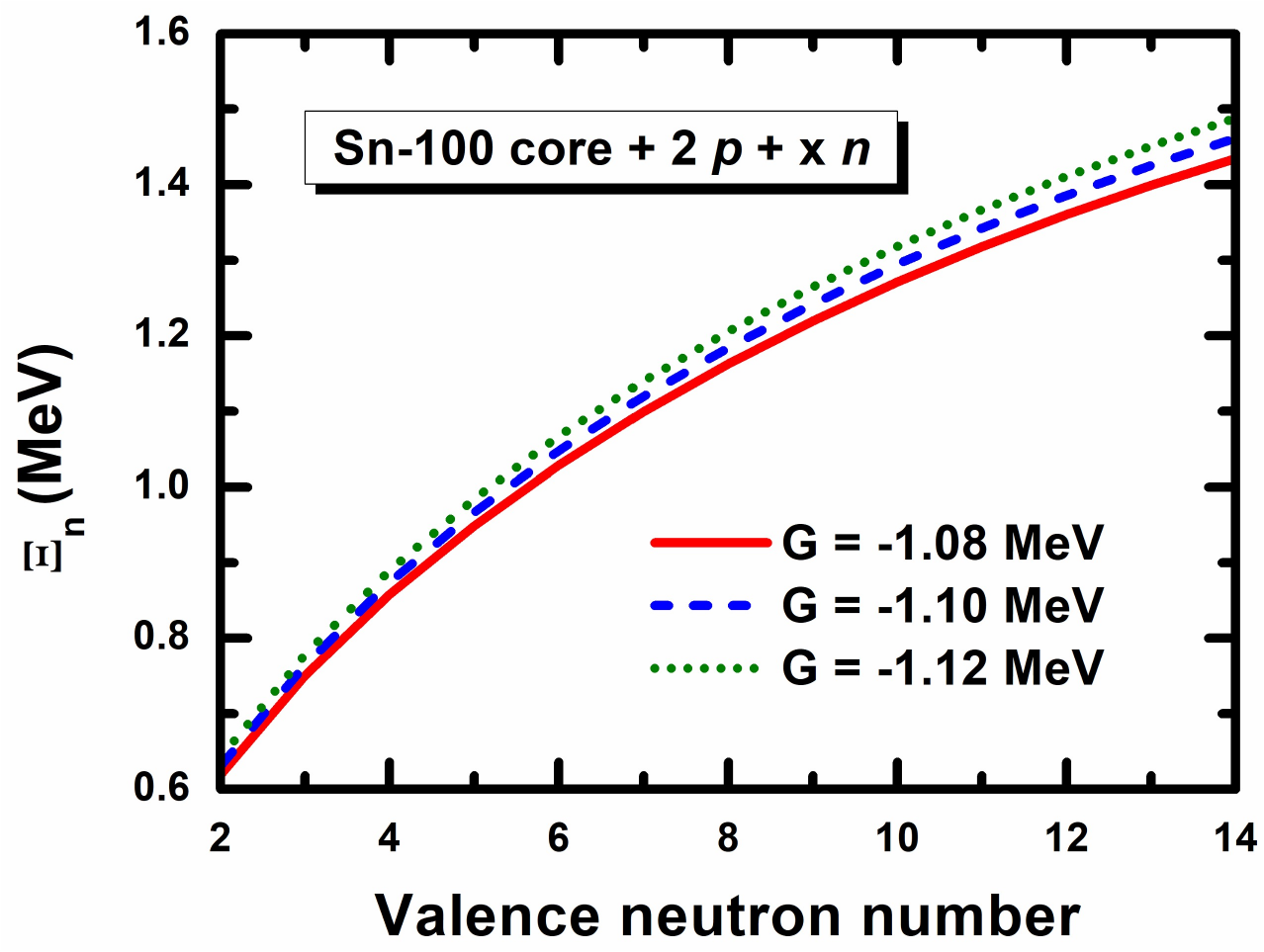}
  \caption{The degeneracy-weighted neutron gap $\Xi_n$ as a function of valence neutron number outside the \(^{100}\)Sn core.}\label{Fig:1}
\end{figure}

In the present study, the isotopic chain from \(^{104}\)Te to
\(^{116}\)Te is specifically investigated.  Anchoring the calibration
to \(^{112}\)Te and \(^{114}\)Te gives \(G\simeq -1.10~\mathrm{MeV}\),
for which
\begin{equation}
\Xi_n(^{112}\mathrm{Te}) \approx 1.30~\mathrm{MeV},
\qquad
\Xi_n(^{114}\mathrm{Te}) \approx 1.39~\mathrm{MeV},
\end{equation}
which lie within the empirical interval.
In Fig.~\ref{Fig:1}, the degeneracy-weighted neutron gap $\Xi_n$ is shown as a function of valence neutron number out of Sn-$100$ core.
As the reference value in the following calculations, a reasonable uncertainty window of approximately $G=-1.08~\mathrm{MeV}\text{--}-1.12~\mathrm{MeV}$ is also adopted.

\begin{figure}
\includegraphics[width=\columnwidth]{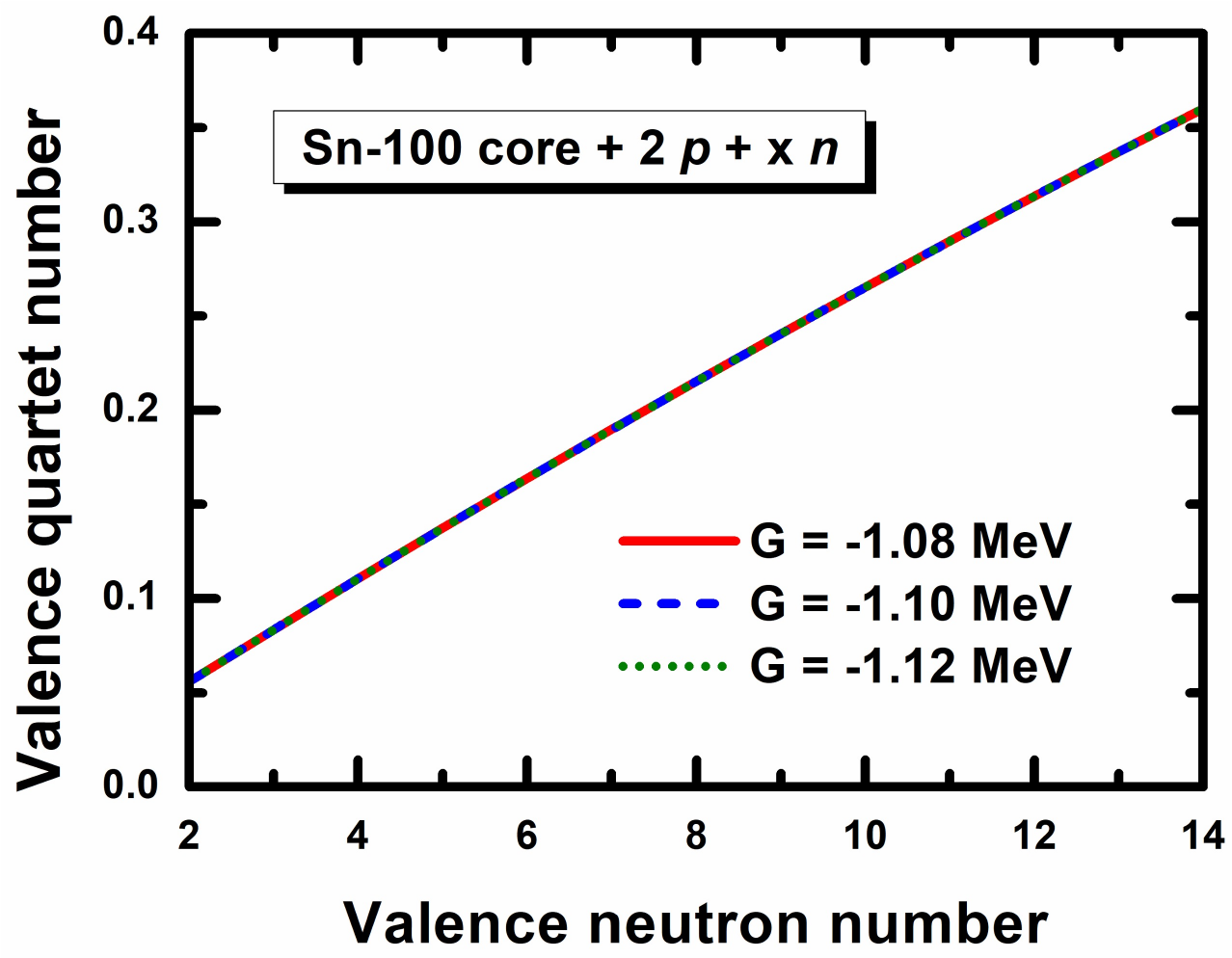}
  \caption{The valence quartet number as a function of valence neutron number out of \(^{100}\)Sn core.}\label{Fig:2}
\end{figure}
In addition, the quartet number is defined as
\begin{equation}
N_Q
=
\sum_b m_bw_b^2,
\end{equation}
where $m_b$ denotes the multiplicity of block $b$.
By scanning over the isotopic chain from \(^{104}\)Te to \(^{116}\)Te, the valence quartet number as a function of valence neutron number out of Sn-$100$ core is given in Fig.~\ref{Fig:2}.
As the neutron number increases at fixed \(Z=2\), the calculated \(N_Q\) increases, indicating that the quartet component indeed becomes larger. 
An increasing trend of \(N_Q\) is obtained over the whole range of valence neutron numbers considered here.  This behavior shows that, within the present quartet BCS variational space, the additional neutrons do not merely fill independent neutron-pair configurations, but also enhance the admixture of the quartet component in the correlated ground state.  
Since the number of valence protons is fixed to \(Z=2\), the growth of \(N_Q\) should be understood as an increasing probability for the two valence protons to participate in neutron-proton correlated four-body configurations as more valence neutrons become available.

It is also worth noting that the three curves obtained with
\(G=-1.08\), \(-1.10\), and \(-1.12\) MeV almost overlap.  Thus, in the
present parameter range, the predicted quartet number is rather insensitive
to the moderate uncertainty in the pairing strength.  The dominant
dependence of \(N_Q\) is instead controlled by the valence neutron number.

We have also checked the corresponding condensed quartet number~\cite{Guo2022Phys.Rev.C105.024317,Guo2025Phys.Rev.C112.024310}, defined as
\begin{equation}
N_Q^{\rm cond}
=
\sum_b m_b u_b^2 w_b^2 .
\end{equation}
This quantity measures the coherent quartet component associated with the
mixing between the empty-block amplitude \(u_b\) and the quartet amplitude
\(w_b\).  Although it is not shown separately, \(N_Q^{\rm cond}\) exhibits
the same monotonic increasing behavior as \(N_Q\) along the isotopic chain.
This provides an additional indication that the increasing $N_Q$ corresponds to an enhanced coherent quartet component within the present quartet BCS ansatz, rather than only to a redistribution of normalization among the variational amplitudes.
One may also construct pair-quartet mixing measures, for example
\(\sum_b m_b(\sum_{T_3}|v_{b,T_3}|^2)|w_b|^2\), to characterize the coexistence of pair-like and quartet-like components within the same block.  
Here we use \(N_Q^{\rm cond}\) only as an analog of the BCS
factor \(u^2v^2\), isolating the coherent empty-block--quartet mixing.

\begin{figure}
\includegraphics[width=\columnwidth]{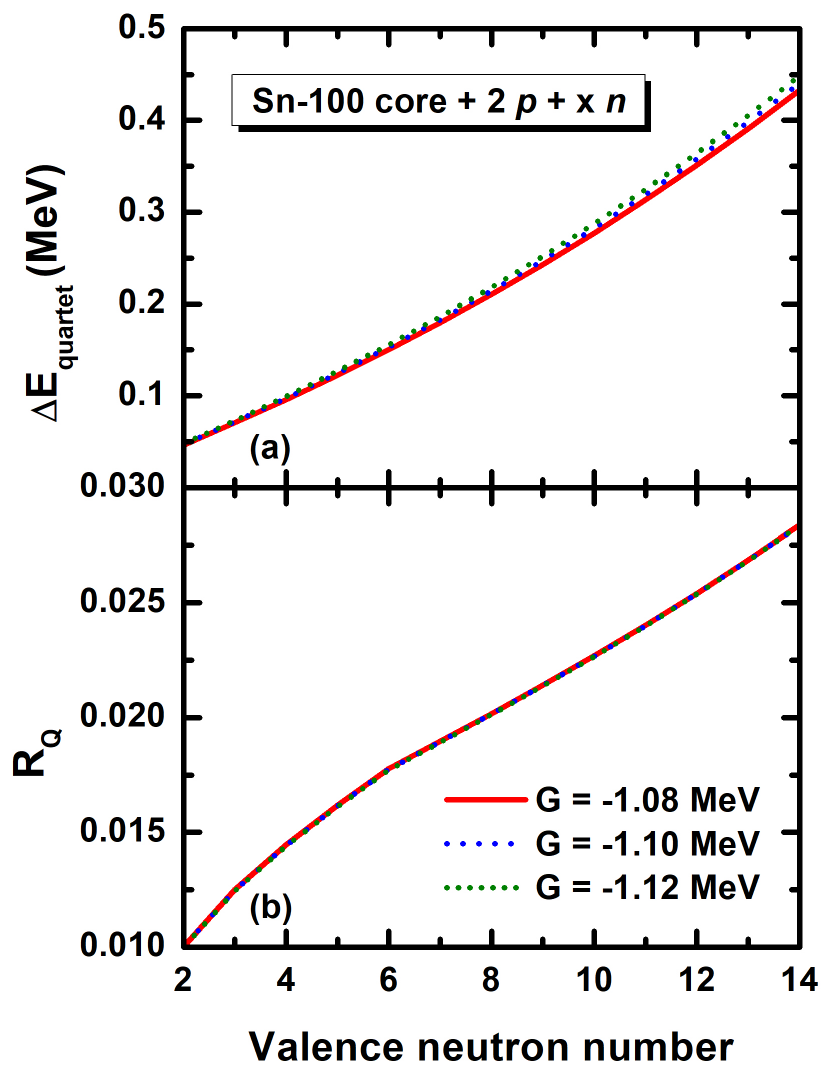}
  \caption{(a) Quartet-induced energy gain
\(\Delta E_{\rm quartet}=E_{\rm noQ}^{\rm qBCS}-E_{\rm full}^{\rm qBCS}\)
as a function of valence neutron number out of the \(^{100}\)Sn core.
(b) Corresponding fraction
\(R_Q=\Delta E_{\rm quartet}/E_{\rm corr}\), where
\(E_{\rm corr}=E_{\rm ref}-E_{\rm full}^{\rm qBCS}\) and
\(E_{\rm ref}\) is the uncorrelated filling energy in the same valence
space.}\label{Fig:3}
\end{figure}

To quantify the energetic effect of the explicit quartet amplitude, we define the quartet-induced energy gain as
\begin{align}
\Delta E_{\rm quartet}
=
E_{\rm noQ}^{\rm qBCS}
-
E_{\rm full}^{\rm qBCS}.
\end{align}
Here, \(E_{\rm full}^{\rm qBCS}\) is the expectation value of the Hamiltonian obtained from the full quartet BCS variational state, while 
\(E_{\rm noQ}^{\rm qBCS}\) is obtained by constraining the quartet  amplitude to zero, namely \(w_b=0\), but retainkeeping the two-pair block structure of Eq.~\eqref{twf}.
Consequently, the resulting state does not reduce to the direct product of independent neutron and proton BCS states, and same quartet BCS block structure.  
Therefore, \(\Delta E_{\rm quartet}\) should be interpreted as the energy gain associated specifically with allowing the explicit quartet amplitude within the quartet BCS ansatz.  
It is not the energy difference between the full quartet BCS state and the conventional factorized BCS state because of the $T=1$ $np$ pair in the trial wave function~\eqref{twf}.

To estimate the importance of this energy gain relative to the total
correlation energy, we also introduce the uncorrelated reference energy
\(E_{\rm ref}\), obtained by filling the \(2d_{5/2}\) and \(1g_{7/2}\)
single-particle levels with the same valence neutron and proton numbers,
but without pairing or quartet correlations.  We then define
\begin{align}
E_{\rm corr}
&=
E_{\rm ref}
-
E_{\rm full}^{\rm qBCS},
&
R_Q
&=
\frac{\Delta E_{\rm quartet}}{E_{\rm corr}} .
\end{align}

The resulting \(\Delta E_{\rm quartet}\) is shown in Fig.~\ref{Fig:3}(a).  
A positive value indicates that the inclusion of the quartet amplitude lowers the variational energy and hence provides additional binding within the present model space.  
The energy gain increases monotonically with valence neutron number, from about \(0.05\) MeV near \(N_{\rm val}=2\) to about \(0.4\)--\(0.45\) MeV near \(N_{\rm val}=14\).  This trend is consistent with the growth of the quartet number shown in Fig.~\ref{Fig:2}: as more valence neutrons are added, the quartet component becomes larger and its
contribution to the variational energy becomes more important.

Figure~\ref{Fig:3}(b) shows the corresponding ratio \(R_Q\). 
The ratio also increases with valence neutron number, from about \(1\%\) to about \(2.8\%\) over the isotopic chain considered here.  Thus, the explicit quartet amplitude gives a steadily growing extra binding contribution, although it remains a small fraction of the total correlation energy measured relative to the uncorrelated filling reference.

{The absolute quartet measures obtained here cannot be compared directly with the quartet components in the particle-number-projected calculations~\cite{Chasman2002PhysLettB524.81,Chasman2003PhysLettB553.204}, because the trial spaces and the quantities used to characterize quartet correlations differ.  
Those calculations focused on systems at or close to $N=Z$ and allowed both $T=1$ and $T=0$ neutron-proton pairing channels.  
In the present calculation, $Z_{\rm val}=2$ specifies the average number of protons outside the heavy $^{100}$Sn core.  
This small active-proton number restricts the proton phase space available for quartet correlations, but the relevant single-particle spectrum and interactions remain those of the $^{100}$Sn region.  
The small average number of active protons is an important factor limiting the quartet admixture.  
In addition, the calculations also differ in neutron-proton asymmetry, single-particle configurations and spacings, pairing channels, and particle-number treatment.  
In particular, an average active-proton number of two does not by itself exclude a substantial one-quartet admixture when the active orbits and interactions are favorable.  
Within the present model, the more robust result is the monotonic neutron-number dependence obtained under the same Hamiltonian and model-space assumptions, whereas the absolute quartet fraction remains model dependent.}

The dependence on the pairing strength is also modest in the range \(G=-1.08\) to \(-1.12\)~MeV.  
A slightly stronger attraction gives a slightly larger absolute energy gain, as expected, while the ratio \(R_Q\) changes only weakly.  
Thus, within the parameter range considered here, the increasing trends of \(\Delta E_{\rm quartet}\) and \(R_Q\) are insensitive to the moderate variation of \(G\).

\begin{figure}
\includegraphics[width=\columnwidth]{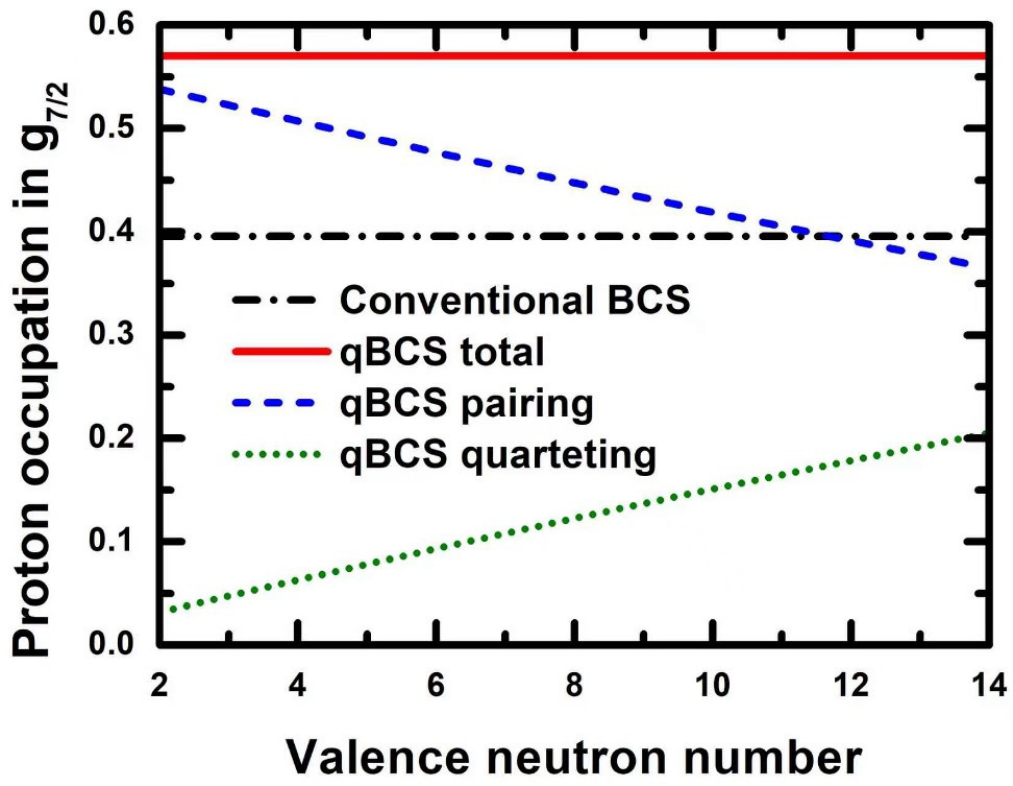}
\caption{Proton occupation fraction of the \(1g_{7/2}\) orbit as a function of valence neutron number out of the \(^{100}\)Sn core.  
The dash-dotted line denotes the conventional BCS result.  
The solid line shows the total quartet BCS occupation, while the dashed and dotted lines show its pair-like and quartet-induced contributions, respectively.
The pair-like and quartet-induced contributions are defined in Eqs.~\eqref{eq:proton-pair-occupation} and \eqref{eq:proton-quartet-occupation}, respectively.
}\label{Fig:4}
\end{figure}

In order to further clarify the microscopic origin of the enhanced quartet correlations, Fig.~\ref{Fig:4} shows the proton occupation  
\begin{align}
P_{\pi,g}=
\frac{N_{\pi,g}}{Z}
\end{align}
of the \(1g_{7/2}\) orbit.
For the full quartet BCS state it is decomposed as
\begin{align}
P_{\pi,g}^{\rm qBCS}
=
P_{\pi,g}^{\rm pair}
+
P_{\pi,g}^{\rm quartet},
\end{align}
where the numbers of protons involved in the pairing and quarteting are given as
\begin{align}\label{eq:proton-pair-occupation}
P_{\pi,g}^{\rm pair}
=
\frac{1}{Z}
\sum_b m_b v_{b,-1}^2
\end{align}
and
\begin{align}\label{eq:proton-quartet-occupation}
P_{\pi,g}^{\rm quartet}
=
\frac{1}{Z}
\sum_b m_b w_b^2,
\end{align}
respectively.

In the conventional BCS reference state, the
neutron and proton condensates are factorized, and only like-particle
pairing is included.  Consequently, the proton occupation is mainly
controlled by the proton single-particle spectrum.  Since the
\(2d_{5/2}\) orbit is lower in energy than the \(1g_{7/2}\) orbit, the
conventional BCS calculation gives a relatively smaller \(1g_{7/2}\)
occupation.
In contrast, the full quartet BCS state contains neutron-pair, proton-pair, and quartet amplitudes within the same variational block.  
This correlated block structure allows self-consistent feedback between the neutron and proton sectors and strongly mixes the \(2d_{5/2}\) and \(1g_{7/2}\) orbits.
As a result, the total quartet BCS proton occupation of \(1g_{7/2}\) is already close to the degeneracy-weighted limit,
\begin{align}
\frac{\Omega_g}{\Omega_d+\Omega_g}
=
\frac{4}{7},
\end{align}
even at small valence neutron number. 
This indicates that quartet correlations largely overcome the single-particle preference for the lower \(2d_{5/2}\) orbit and make use of the larger \(1g_{7/2}\) degeneracy.

Although the total quartet BCS \(1g_{7/2}\) occupation changes only weakly with neutron number, its internal composition changes significantly.  
The pair-like contribution decreases as the valence neutron number increases, whereas the quartet contribution increases almost linearly.  
This behavior reflects the fixed proton-number constraint: as more proton weight is carried by quartet configurations, the pure proton-pair component is reduced.  
Therefore, the increase of the quartet number with neutron number is not primarily driven by a further increase of the total \(1g_{7/2}\) proton occupation, but by a redistribution of the proton weight from pair-like configurations to quartet configurations within an already strongly mixed \(d_{5/2}\oplus g_{7/2}\) space.

\begin{figure}
\includegraphics[width=\columnwidth]{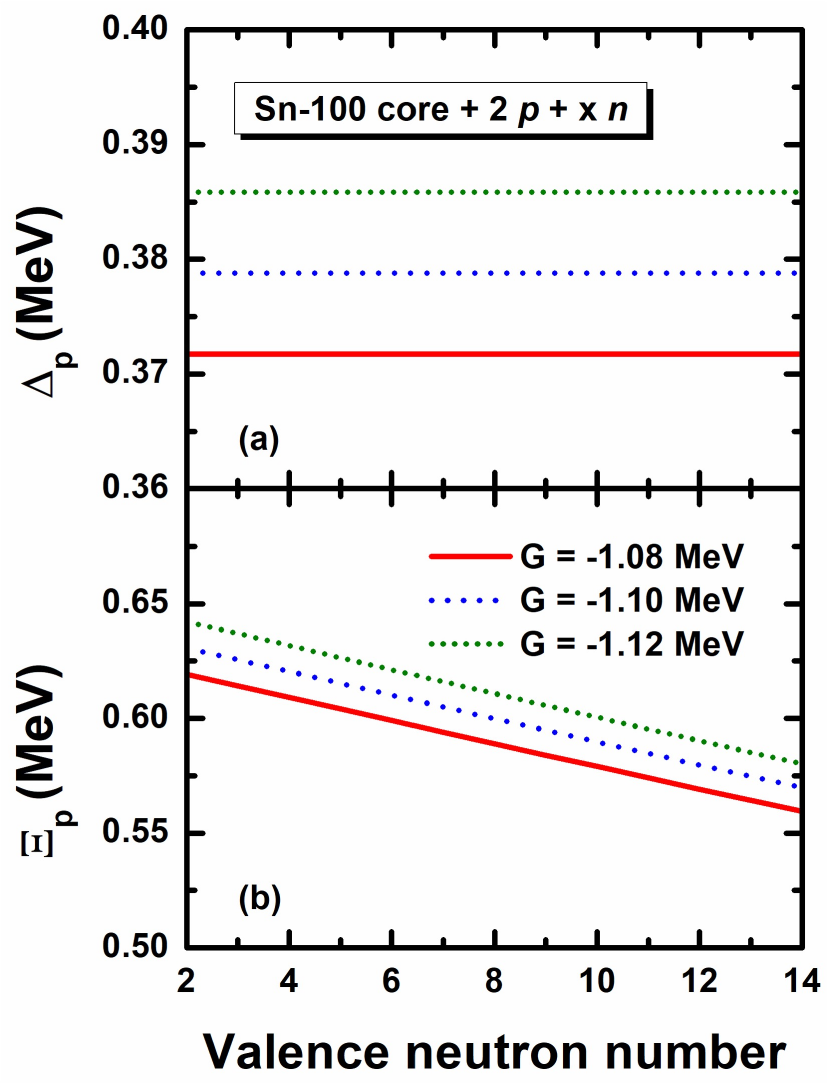}
\caption{Proton pairing quantities as functions of valence neutron number
out of the \(^{100}\)Sn core.  
(a)~Conventional BCS proton
gap \(\Delta_p\), obtained from the factorized like-particle BCS reference.
(b)~Quartet BCS proton gap-like quantity \(\Xi_p\).  
The three curves correspond to \(G=-1.08\), \(-1.10\), and \(-1.12\)~MeV.}\label{Fig:5}
\end{figure}

Figure~\ref{Fig:5} compares the proton pairing field in the conventional BCS reference with the corresponding proton gap-like quantity in the full quartet BCS calculation.  
In the conventional BCS reference state, the total wave function is written
as a direct product of independent neutron and proton BCS states,
\begin{align}
|\Psi_{\rm BCS}\rangle=|\Psi_n\rangle\otimes|\Psi_p\rangle .
\end{align}
As a result, at fixed \(Z=2\), fixed proton single-particle energies, and
fixed \(G\), the proton gap has no explicit dependence on the valence
neutron number.
Therefore, at fixed valence proton number \(Z=2\), fixed proton single-particle energies, and fixed pairing strength \(G\), the proton gap \(\Delta_p\) is independent of the valence neutron number.  
This gives the almost horizontal curves in Fig.~\ref{Fig:5}(a).

In contrast, the quartet BCS quantity shown in Fig.~\ref{Fig:5}(b),
\begin{align}
\Xi_p
\equiv
{\Xi_{-1}}
=
\frac{3\Delta_{d,-1}+4\Delta_{g,-1}}{7},
\end{align}
should be interpreted as an effective proton-pairing field in the correlated quartet BCS state, rather than as the conventional BCS gap.  
In the quartet BCS wave function, this field contains both the ordinary proton-pair coherence and the quartet-induced proton-pair transfer coherence,
schematically
\begin{align}
\Xi_{-1}
\propto
\sum_b m_b
\left(
{\mathcal A}_b u_bv_{b,-1}
+
{\mathcal B}_b w_bv_{b,+1}
\right).
\end{align}
Although the quartet-induced term increases with neutron number, the fixed proton-number constraint causes the pure proton-pair coherence
\(u_bv_{b,-1}\) to decrease as more proton weight is carried by the quartet component.  
Since the pair-like term is the dominant contribution, the
total quartet BCS proton gap-like quantity \(\Xi_p\) decreases gradually with
valence neutron number.

The quartet BCS values remain larger than the conventional BCS gaps over the whole isotopic chain.  
This reflects the additional neutron-proton feedback and pair-field dressing generated by the correlated quartet BCS block structure.  
At the same time, the downward trend of \(\Xi_p\) is consistent with the occupation analysis in Fig.~\ref{Fig:4}, where increasing neutron number mainly redistributes the fixed proton weight from the pair-like component to the quartet component, thereby reducing the ordinary proton-pair coherence.

\section{Summary and perspectives}\label{sec:IV}

We have investigated intrinsic quartet correlations in neutron-rich Te isotopes within a finite-nucleus quartet BCS framework.  
Taking $^{100}$Sn as an inert core, we considered two valence protons and valence neutrons in the $2d_{5/2}\oplus1g_{7/2}$ model space and used a charge-independent isovector $J=0$, $T=1$ pairing interaction.  
The effective pairing strength was fixed from empirical neutron pairing gaps in Te isotopes away from the proton dripline and then used to extrapolate toward the more proton-rich side.

The main result is that adding valence neutrons at fixed valence proton number enhances the quartet component in the correlated quartet BCS state.  
This appears consistently in the quartet number, the condensed quartet number, and the quartet-induced energy gain, and is accompanied by a redistribution of the fixed proton weight from pair-like configurations to quartet-like configurations.  
To this end, within the present variational space, the additional neutrons do not act only as spectators filling neutron-pair states, but also modify the neutron-proton four-body correlations.

{The present calculation complements earlier studies of quartet and neutron-proton pairing~\cite{Chasman2002PhysLettB524.81,Chasman2003PhysLettB553.204,Chasman2003PhysLettB577.47,Chasman2007PhysRevLett99.082501,Baran2020Phys.Lett.B805.135462,Baran2020Phys.Rev.C102.061301,Guo2022Phys.Rev.C105.024317} by applying the two-pair-block quartet BCS ansatz of Eq.~\eqref{twf} to the neutron-number dependence of quartet correlations in neutron-rich Te isotopes.  
Within the adopted model space, the quartet number, condensed quartet number, and quartet-induced energy gain all increase consistently with neutron number, providing a qualitative indication that additional valence neutrons enhance quartet correlations.  
Extending the analysis to wider isotopic and isotonic chains will allow the roles of shell structure, neutron excess, and weak binding to be assessed more systematically.}

{The intrinsic quartet number $N_Q$ calculated here is an internal measure of four-body correlations rather than an $\alpha$-cluster preformation probability or a reaction observable~\cite{Tanaka2021Science371.260264,Nakatsukasa2023Phys.Rev.C108.014318,Yoshida2026Prog.Theor.Exp.Phys.2026.04A108}. 
Its monotonic increase may nevertheless provide a useful model-dependent trend whose robustness can be examined in enlarged valence spaces.  
Establishing a quantitative connection with $\alpha$-emission, transfer, or knockout observables will require isoscalar neutron-proton correlations, particle-number restoration, continuum coupling, deformation, and an explicit reaction treatment.}

\begin{acknowledgments}
The authors thank Youngman Kim, Jian Li, Haozhao Liang, and Hiroyuki Tajima for useful discussions.
This work is supported in part by the National Research Foundation of Korea (NRF), funded by Ministry of Science and ICT (RS-2024-00436392), and by the RIKEN TRIP initiative (Nuclear Transmutation).
Y.G. is supported by RIKEN Special Postdoctoral Researchers Program.
H.S. is supported by the JSPS Grant-in-Aid for Scientific Research (C) under Grant No. JP26K07079.
\end{acknowledgments}


%

\end{CJK}
\end{document}